\documentclass[prl,twocolumn,amsmath,amssymb,showpacs,superscriptaddress]{revtex4}%mine
\usepackage{graphicx}% Include figure files
\usepackage{dcolumn}% Align table columns on decimal point
\usepackage{bm}% bold math
\usepackage{psfrag}%not using?
\def\be{\begin{equation}}
\def\ee{\end{equation}}
\def\ba{\begin{eqnarray}}
\def\ea{\end{eqnarray}}
\newcommand{\ie}{\emph{i.e.\ }}
\newcommand{\eg}{\emph{e.g.\ }}
\newcommand{\ket}[1]{|#1 \rangle}

\newcommand{\Ham}{\mathcal{H}}
\newcommand{\energy}{\mathcal{E}}

\begin{document}

%\preprint{APS/123-QED}

% Force line breaks with \\
%\title{Tree-Level Electron-Photon Interactions in Graphene}
%\title{Interactions between Graphene Electrons and Three-Dimensional Photons}
\title{Spin and the Honeycomb Lattice: Lessons from Graphene}

\author{Matthew Mecklenburg}
%\email{meck0005@physics.ucla.edu}
\affiliation{Department of Physics and Astronomy, University of California, Los Angeles, California 90095, U.S.A.}
\affiliation{California NanoSystems Institute, University of California, Los Angeles, California 90095, U.S.A.}

\author{B.C. Regan}
\email{regan@physics.ucla.edu}
\affiliation{Department of Physics and Astronomy, University of California, Los Angeles, California 90095, U.S.A.}
\affiliation{California NanoSystems Institute, University of California, Los Angeles, California 90095, U.S.A.}


\date{October 31, 2010}
             %  but any date may be explicitly specified February 9, 2010

\begin{abstract}
Spin-1/2 particles such as the electron are described by the Dirac equation, which allows for two spin eigenvalues (up or down) and two types of energy eigenvalues (positive or negative, corresponding to the electron and the positron).  A model of electrons hopping from atom to atom in graphene's honeycomb lattice gives low-energy electronic excitations that obey a relation formally identical to a $2+1$ dimensional Dirac equation.   Graphene's spin equivalent, ``pseudospin'', arises from the degeneracy introduced by the honeycomb lattice's two inequivalent atomic sites per unit cell.   Previously it has been thought that the usual electron spin and the pseudospin indexing the graphene sublattice state are merely analogues.  Here we show that the pseudospin is also a real angular momentum.   This identification explains the suppression of electron backscattering in carbon nanotubes and the angular dependence of light absorption by graphene. Furthermore, it demonstrates that half-integer spin like that carried by the quarks and leptons can derive from hidden substructure, not of the particles themselves, but rather of the space in which these particles live.% Furthermore, it demonstrates that experimentally accessible half-integer spin can arise as a result of quantizing space with a non-trivial lattice.

\end{abstract}

%\pacs{78.67.Wj, 78.67.Ch, 13.40.Hq}
%78.67.Wj 	Optical properties of graphene 
%78.67.Ch Optical properties of low-dimensional, mesoscopic, and nanoscale materials and structures,nanotubes
%13.40.Hq 	Electromagnetic decays
%73.22.Pr 	Electronic structure of graphene 
%03.65.Pm 	Relativistic wave equations (listed as Dirac equation in alphabetical index)
%71.10.Fd 	Lattice fermion models (Hubbard model, etc.) ... don't like the parenthetical
%11.15.Ha Lattice gauge theory
%11.30.-j 	Symmetry and conservation laws
%unbelievable!  Nothing in PACS on spin or origins thereof.
%Lattice theory and statistics, 05.50.+q  in 05. 	Statistical physics, thermodynamics, and nonlinear dynamical systems
%04.60.Nc 	Lattice and discrete methods in 	Quantum gravity in 04. 	General relativity and gravitation 
%Borici PACS numbers: 11.15.Ha Lattice gauge theory, 11.30.Rd Chiral symmetries , 71.10.Fd, 71.20.-b Electron density of states and band structure of crystalline solids

\maketitle

``Spin'' refers to an angular momentum that has no classical analogue --- it is not possible to understand spin in terms of a mechanical model of some rotating object \cite{1997Tomonaga}.  The net angular momentum of composite particles, such as protons, neutrons, atoms, and molecules, derives from the spins of their constituents, plus any orbital angular momenta due to the constituents' relative motion.  Spin and orbital angular momenta are quantitatively distinguishable, since the former can be half-integer while the latter take on integer values only, measured in units of the reduced Planck constant $\hbar$.  In the standard model of particle physics the ultimate constituents of matter are the quarks and leptons.  These particles have no internal structure down to length scales of $10^{-18}$~m (limited by the collision energies $\sim 200$~G$e$V currently achievable in particle accelerators) \cite{2008PDGcompositeness}, so their spins are considered intrinsic.  

The deepest insight into the origin of spin has been provided by Dirac, who manipulated Einstein's quadratic energy-momentum relation $E^2=p^2 c^2 + m^2 c^4$ to give a linear equation consistent with the postulates of quantum mechanics \cite{1958Dirac}.  Dirac's equation predicts not only spin but also antiparticles, which were unknown at the time.  Thus we understand, for example, the electron's spin $\hbar/2$ and the existence of the positron as natural consequences of the Dirac equation, which is built on the theories of relativity and quantum mechanics.  As was pointed out 25 years ago \cite{1984DiVincenzo,1984Semenoff}, the low-energy electronic excitations in graphene obey a $2+1$ dimensional Dirac equation, with holes and the sublattice state playing the role of positrons and spin respectively.  In this Letter we show that the sublattice state vector describes an `intrinsic' angular momentum in $3+1$ dimensions.  This identification provides a physical model that associates spin with an underlying structure.  Unlike the case of composite particles, where spin follows from other spins, in this example the spin $\hbar/2$ is a consequence of the nontrivial spatial lattice, invisible at low energies, that hosts the particle.

Graphene's Dirac equation follows from the tight-binding (\ie hopping) model, which was first applied to graphene by Wallace \cite{1947Wallace}. With the experimental isolation of carbon nanotubes \cite{1991Iijima} and, more recently, graphene itself \cite{2004Novoselov}, the hopping model has been shown to give an effective description of these real materials \cite{1998Saito,2009NetoReview}.  Furthermore, graphene equivalents of the quantum relativistic effects implied by the Dirac equation, such as  Klein tunneling and \emph{Zitterbewegung}, are now experimentally accessible, making this condensed matter system a practical test bed for these particle physics phenomena \cite{2007KatsnelsonBridge,2006KatsnelsonKlein}.  %the contents of \cite{2006KatsnelsonZitterbewegung} is  included in Bridge

\begin{figure}\begin{center}
	\includegraphics[width=0.9\columnwidth]{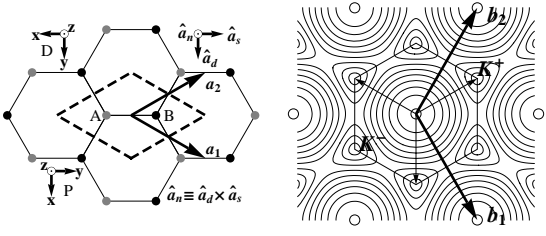}%{spingeo.eps}3.25in
	\caption{\label{fig:geo} The honeycomb lattice in real space (left) and the corresponding band structure in reciprocal space (right).  On the left, gray and black circles represent A and B lattice sites respectively.  One choice of unit cell is demarcated with a dashed line.  Coordinate axes oriented to give the Pauli and Dirac forms of the electronic Hamiltonian are labeled ``P'' and ``D''. On the right, the hexagonal first Brillouin zone is shown, with the positions of $\mathbf{K}^+$ points indicated by thin arrows.  While generally the contours of constant energy [based on the full Hamiltonian (\ref{eq:Mmatrix})] show threefold or sixfold rotational symmetry, near the points $\mathbf{K}^{\pm}$ they are circular \cite{1998Saito,2009Bena}.}
\end{center}\end{figure}

We first show how the hopping model produces the 2+1 dimensional Dirac equation, working in 3+1 dimensions but without initial reference to specific coordinate axes.  Graphene's electronic states are described as a linear combination of atomic orbitals constructed to fulfill the Bloch condition $\Psi_{\mathbf{Q}}(\mathbf{r}+\mathbf{R})= e^{i \mathbf{Q}\cdot \mathbf{R}}\Psi_{\mathbf{Q}}(\mathbf{r})$, 
\begin{equation}\label{eq:wavefunction}
\Psi_{\mathbf{Q}}(\mathbf{r})=\sum_j^N \frac{e^{i \mathbf{Q}\cdot \mathbf{R_j}}}{\sqrt{N}}\left[c^A_{\mathbf{Q}}\phi(\mathbf{r}-\mathbf{R}^A_j)+c^B_{\mathbf{Q}}\phi(\mathbf{r}-\mathbf{R}^B_j)\right],
\end{equation}
where $j=(m,n)$ indexes the $N$ sites of the hexagonal Bravais lattice described by $\mathbf{R}_j=m \mathbf{a}_1+n \mathbf{a}_2$, and the vectors $\mathbf{R}^{A}_j$ ($\mathbf{R}^{B}_j$) point to the `A' (`B') sublattice sites within the unit cell $j$, respectively (see Fig.~\ref{fig:geo}).  Indices labeling the usual electron spin have been suppressed for notational convenience and will be henceforth.  The coefficients $c_{\mathbf{Q}}$ multiplying the carbon atoms' $2P_z$ atomic orbitals $\phi(\mathbf{r})$ are chosen to solve the crystal Hamiltonian.  In the literature there are equivalent alternatives \cite{1947Wallace,1984Semenoff,2009NetoReview} to the expansion (\ref{eq:wavefunction}), but this choice reveals the momentum space symmetry of the Hamiltonian more readily \cite{2009Bena}.

The electronic Hamiltonian contains two kinds of terms, one representing an electron's energy on a particular site, and the other representing the energy advantage conferred by the freedom to hop to a neighboring site. The amplitude for nearest-neighbor hopping is parametrized by an energy $t$, and  next-nearest-neighbor and higher order effects are neglected. An electron occupying the state $\phi(\mathbf{r}-\mathbf{R}^A_j)$ is described by an operator $A^\dagger_{\mathbf{R}_j}$ that creates an electron on the ``A'' site in cell $j$ when it acts on the vacuum state $\ket{0}$.  With similar language for the ``B'' sites, the graphene Hamiltonian is
\begin{equation}\label{eq:originalTB}
H =  \sum_{j} (\energy_A A_{\mathbf{R_j}}^{\dagger} A_{\mathbf{R_j}}^{\phantom{\dagger}} +  \energy_B B^{\dagger}_{\mathbf{R_j}} B^{\phantom{\dagger}}_{\mathbf{R_j}})% \nonumber \\&\qquad
-t \sum_{<i,j>} (A^{\dagger}_{\mathbf{R_i}} B^{\phantom{\dagger}}_{\mathbf{R_j}} + \textrm{h.c.}),
\end{equation}
where the notation $\langle i,j\rangle$ indicates sums over all sites $j$ and their nearest neighbors $i$.  In graphene the A sites are identical to the B sites modulo a $\pi$ rotation, so the energies $\energy_A=\energy_B$.  Following Semenoff \cite{1984Semenoff}, we allow for different site energies ($\energy_A\neq \energy_B$), as in hexagonal boron nitride, and take the graphene limit where appropriate.

%\vspace{-1.2in}
Introducing the Fourier transform of the operators $A_{\mathbf{R}_i}=\sum_{j} A_{\mathbf{Q}_j} \exp(i \mathbf{R}_i\cdot\mathbf{Q}_j)/\sqrt{N}$, where the $\mathbf{Q}_j=\frac{m}{N_1} \mathbf{b}_1+\frac{n}{N_2} \mathbf{b}_2$ are the $N=N_1 N_2$ wave vectors in the first Brillouin zone, puts the Hamiltonian (\ref{eq:originalTB}) in the form
\begin{equation}\label{eq:fourierTB}
H =  \sum_{j} \begin{pmatrix} A^{\dagger}_{\mathbf{Q}_j} & B^{\dagger}_{\mathbf{Q}_j} \end{pmatrix} \Ham
\begin{pmatrix} A^{\phantom{\dagger}}_{\mathbf{Q}_j} \\ B^{\phantom{\dagger}}_{\mathbf{Q}_j} \end{pmatrix},
\end{equation}
where we have used the closure relation $\sum_{j} e^{i \mathbf{R}_j\cdot (\mathbf{Q}-\mathbf{Q}')}= N \delta_{\mathbf{Q},\mathbf{Q}'}$, and defined the single particle Hamiltonian
\begin{equation}\label{eq:Mmatrix}
\Ham= -t\begin{pmatrix}
-\Delta/t &1+ e^{-i \mathbf{Q}\cdot \mathbf{a}_1}+e^{-i \mathbf{Q}\cdot \mathbf{a}_2}\\
1+ e^{i \mathbf{Q}\cdot \mathbf{a}_1}+e^{i \mathbf{Q}\cdot \mathbf{a}_2}  & \Delta/t
\end{pmatrix}.
\end{equation}
Here the energy difference $\Delta$ is defined by $\Delta\equiv(\energy_A-\energy_B)/2$, and the energy origin is chosen such that $(\energy_A+\energy_B)/2=0$.  Graphene has two atoms per unit cell, each of which donates one electron to the valence band, so in the lowest approximation the first Brillouin zone is exactly filled.  Thus the Fermi energy is zero here. 

The off-diagonal matrix elements of the $\Ham$ matrix (\ref{eq:Mmatrix}) vanish at the corners $\mathbf{Q} = \mathbf{K}$ of the Brillouin zone (see Fig.~\ref{fig:geo}), which in the graphene case $\energy_A=\energy_B$ leads to the famous degeneracy at these `Dirac points': $\mathbf{K}^\kappa=\kappa \frac{2 \mathbf{b}_2+\mathbf{b}_1}{3}+m \mathbf{b}_1+n \mathbf{b}_2$ \cite{2009Bena}.  Here $\kappa=\pm1$ is the ``valley'' index that distinguishes the two inequivalent types of $\mathbf{K}$ points.  To see the structure of low-energy excitations near the Dirac points we define $\mathbf{k}=\mathbf{Q}-\mathbf{K}$ and restrict our analysis to the case where $\mathbf{k} \cdot \mathbf{a}_i$ is small. Recalling $\mathbf{a}_i\cdot\mathbf{b}_j=2 \pi \delta_{ij}$, to lowest order 
\begin{equation}\label{eq:linearMmatrix}
\Ham=\begin{pmatrix}
\Delta & \sqrt{3} t a \mathbf{k}(\kappa \hat{\mathbf{a}}_d -i \hat{\mathbf{a}}_s)/2 \\
 \sqrt{3} t a \mathbf{k}(\kappa \hat{\mathbf{a}}_d +i \hat{\mathbf{a}}_s)/2  & -\Delta
\end{pmatrix},
\end{equation}
where we have defined difference  and sum unit vectors $\hat{\mathbf{a}}_d=(\mathbf{a}_1-\mathbf{a}_2)/ a$ and $\hat{\mathbf{a}}_s=(\mathbf{a}_1+\mathbf{a}_2)/\sqrt{3} a$. (A third vector $\hat{\mathbf{a}}_n\equiv\hat{\mathbf{a}}_d\times\hat{\mathbf{a}}_s$ normal to the plane will also prove useful.) The Hamiltonian (\ref{eq:linearMmatrix}) is a rotational invariant, depending only on scalars and the scalar products of 3D vectors.  As shown in Fig.~\ref{fig:geo}, the eigenvalues of $\Ham$ are independent of the direction of $\mathbf{k}$ near $\mathbf{K}$, which is to say that the Hamiltonian is effectively isotropic in the plane --- the lattice has disappeared.

The analysis leading to (\ref{eq:linearMmatrix}) has been entirely in terms of the basis vectors of the direct and reciprocal lattices of the honeycomb structure, and is independent of any choice of orientation for a 3D Cartesian coordinate system.  The predictions of the theory are independent of this choice.  Two particular orientations put $\Ham$ in familiar and suggestive forms.  Choosing the coordinate orientations labeled ``D'' and ``P'' in Fig.~\ref{fig:geo}, defining the Fermi velocity $v_F=\sqrt{3}a t/2\hbar$, and writing the conjugate momentum $\mathbf{p}=\hbar \mathbf{k}$ gives
\begin{subequations}\label{eq:DiracM}
\begin{align}
 \Ham_D&=v_F(\kappa \sigma_x p_y - \sigma_y p_x )+\sigma_z\Delta \quad \text{and} \label{eq:DiracD}\\%\frac{\sqrt{3}a t}{2\hbar}
\Ham_P&=v_F(\kappa \sigma_x p_x +  \sigma_y p_y) +\sigma_z\Delta ,\label{eq:DiracP}
\end{align}%
\end{subequations}
where the $\sigma_j$ are the usual Pauli matrices. The matrix $\Ham_D$ (\ref{eq:DiracD}) illustrates the connection between the graphene Hamiltonian and the Dirac equation.  In  $3+1$ dimensions the Dirac Hamiltonian is $H=c \gamma^0 (\boldsymbol{\gamma}\cdot \mathbf{p}+mc)$, with $4 \times 4$ matrices $\gamma^\mu=(\gamma^0,\boldsymbol{\gamma})$.  In $2+1$ dimensions the necessary anticommutation relations  $\{\gamma^\mu,\gamma^\nu\}=2 g^{\mu\nu}=2\times \mathrm{diag}(1,-1,-1)$ can be satisfied with a $2\times 2$ representation such as $\gamma^\mu = (\gamma^0,\vec{\gamma}) = (\sigma_z,i \sigma_x, i\kappa \sigma_y)$.  With this definition the Hamiltonian matrix (\ref{eq:DiracD}) becomes $\Ham= v_F \gamma^0(\vec{\gamma}\cdot\vec{p}+ m'v_F)$, with an effective mass $m'$ defined by $\Delta=m'v_F^2$. (We designate the first two components of a 3D vector with an arrow, \eg $\mathbf{p}=p_x \mathbf{\hat{x}}+p_y \mathbf{\hat{y}}+p_z \mathbf{\hat{z}}=\vec{p}+p_z \mathbf{\hat{z}} $.)  Thus this hopping model gives low-energy electronic excitations that obey a $2+1$ dimensional Dirac equation, with the Fermi velocity $v_F$ playing the usual role of the speed of light $c$.%  \cite{1987Griffiths}

With the P coordinate orientation graphene's Hamiltonian matrix has the convenient abbreviation $\Ham=  v_F \vec{\sigma}\cdot \vec{p}\,$ near $\mathbf{K}^+$.  Sometimes this is written $\Ham=  v_F \boldsymbol{\sigma}\cdot \mathbf{p}$, which can give the impression that either $\sigma_z$ or $p_z$ is strictly zero.  In fact, both $\boldsymbol{\sigma}$ and $\mathbf{p}$ have three nonzero components; it just happens that $\sigma_z$ and $p_z$ do not appear in the graphene $\Ham$.   While any state confined to the sheet must have an expectation value \mbox{$\langle p_z\rangle=0$}, because the electrons are localized in $z$ the characteristic magnitude of $p_z$ is $\hbar/a_z$, where $a_z$ is a length scale measuring the $z$ extent of the $2P_z$ orbitals of the expansion (\ref{eq:wavefunction}).  Taking $p_z=0$ is just as improper for an electron in a honeycomb lattice as it is for an electron in an atomic Coulomb potential. The case of $\sigma_z$ will be taken up below. 

It is not obvious whether the operator $\boldsymbol{\sigma}$, which indexes the ``pseudospin'' arising from the $AB$ sublattice degeneracy, corresponds to a real angular momentum \cite{1984DiVincenzo,2007Gusynin}. It might describe a merely analogous two-state system, borrowing the same $SU(2)$ algebra like the isospin symmetry connecting the proton and neutron. Comparison with the 3+1 dimensional Dirac equation makes this second option look likely. The $4\times4$ Dirac matrices give each state two qubits, one each for particle or antiparticle and spin-up or spin-down values. Since the pseudospin labels the band index $\beta$ [see Eq.~\ref{eq:energy} below] and the 2D Dirac matrices are only $2\times2$, it would seem that the one qubit available is engaged.

Surprisingly, this one qubit manages to cover both variables. The pseudospin is related to a real angular momentum, as can be discovered by calculating the commutator of the Hamiltonian with the orbital angular momentum $\mathbf{L}=\mathbf{r}\times \mathbf{p}$ (Ref.~\cite{1958Dirac}),%.  Using the cannonical commutation relation $[r_i,p_j]=i \hbar \delta_{ij}$, we find 
\begin{equation}
[\Ham_P,\mathbf{L}]= - i \hbar v_F 
\begin{pmatrix} 
	\sigma_y p_z \\
 -\kappa \sigma_x p_z\\
  \kappa \sigma_x p_y-\sigma_y p_x 
  \end{pmatrix}.
% =-i\hbar v_F \vec{\sigma}\times \mathbf{p}
\end{equation}
That the Hamiltonian has rotational symmetry about the axis perpendicular to the plane and the commutator $[\Ham_P, L_z]$ is not zero together indicate that there is another angular momentum in the problem.
%table was here
 
In coordinate-independent notation the honeycomb Hamiltonian (\ref{eq:linearMmatrix}) is
\begin{subequations}\label{eq:uAndsig}
\begin{align}
\Ham&= \frac{2 v_F}{\hbar} \mathbf{S} \cdot \mathbf{u},\quad \text{where} \label{eq:HamU}\\
\mathbf{S}&\equiv \frac{\hbar}{2}( \kappa \sigma_x\, \hat{\mathbf{a}}_d + \sigma_y \,\hat{\mathbf{a}}_s + \kappa\sigma_z \,\hat{\mathbf{a}}_n),\quad \text{and} \label{eq:Lspin}\\%\frac{\sqrt{3}a t}{2\hbar}
\mathbf{u}&\equiv (\mathbf{p}\cdot \hat{\mathbf{a}}_d)\,\hat{\mathbf{a}}_d + (\mathbf{p}\cdot \hat{\mathbf{a}}_s)\,\hat{\mathbf{a}}_s +  \kappa(\Delta/v_F)\,  \hat{\mathbf{a}}_n.\label{eq:uCoord}
\end{align}%\boldsymbol{\sigma}
\end{subequations}
The operator $\mathbf{S}$ defined by (\ref{eq:Lspin}) we term ``lattice spin'' to distinguish it from both the dimensionless pseudospin and the usual electron spin.  One can easily verify that $[\Ham, (\mathbf{L}+\mathbf{S})\cdot \hat{\mathbf{a}}_n]=0$ for any value of $\Delta$.  Thus neither the lattice spin nor the orbital angular momentum is separately a constant of the motion, but the projection of the total angular momentum $\mathbf{J}\equiv\mathbf{L}+\mathbf{S}$ onto the plane-normal axis (the $z$-axis for P or D coordinates) is a conserved quantity.  It is not possible to confuse $\mathbf{S}$ with an orbital angular momentum, as its eigenvalues have half-integer magnitude in units of $\hbar$.

In comparison, the connection between pseudospin and angular momentum in a nanotube is less obvious, for in that system the component of the orbital angular momentum along the nanotube axis commutes with the Hamiltonian.  In a carbon ($\Delta=0$) nanotube the component of $\mathbf{S}$ along the axis also commutes with $\Ham$, so the axial projections of the lattice spin and the orbital angular momentum are conserved separately.  Table \ref{DiracTable} summarizes the relevant angular momentum commutation relations for Hamiltonians of the type $H= v\sum^D_i \sigma_i p_i$ with number of spatial dimensions $D=1$, 2, or 3.  

\begin{table}
\begin{tabular}{cccr@{$\times$}lr@{$\times$}lcc}
D&$H$&$\frac{i}{\hbar}[H,\mathbf{r}]$&\multicolumn{2}{c}{$\frac{i}{\hbar}[H,\mathbf{L}]$}&\multicolumn{2}{c}{$\frac{i}{\hbar}[H,\frac{\hbar}{2}\boldsymbol{\sigma}]$}&conserved &system\\
\hline
3&$c \boldsymbol{\sigma}\cdot \mathbf{p}$&$c \boldsymbol{\sigma}$&$ c \boldsymbol{\sigma}$&$ \mathbf{p}$&$ -c \boldsymbol{\sigma}$&$ \mathbf{p}$ & $\mathbf{J}$&  neutrino\\
2&$v \vec{\sigma}\cdot \vec{p}$&$v\vec{\sigma}$& $ v \vec{\sigma}$&$ \mathbf{p}$& $-v \boldsymbol{\sigma}$&$ \vec{p}$& $J_z$& graphene\\
1&$v \sigma_z p_z$&$v\sigma_z$& $ v \sigma_z \mathbf{\hat{z}}$&$ \mathbf{p}$& $ -v \boldsymbol{\sigma}$&$ p_z \mathbf{\hat{z}}$ & $L_z$, $S_z$ & nanotube\\
\end{tabular}%oldschool $ v_F (\boldsymbol{\sigma}-\sigma_z \mathbf{\hat{z}})\times \mathbf{p}$
\caption{Commutation relations for massless Dirac Hamiltonians $H$. In the Heisenberg representation an operator $\mathbf{A}$ without explicit time dependence obeys the equation of motion $d\mathbf{A}/dt=\frac{i}{\hbar}[H,\mathbf{A}]$. The system shorthand refers to right-handed massless neutrinos, graphene in the $xy$ plane, and a metallic carbon nanotube with its axis along the $\mathbf{\hat{z}}$ direction respectively.}
\label{DiracTable}%  We designate the first two components of a 3D vector with an arrow, \eg $\mathbf{p}=p_x \mathbf{\hat{x}}+p_y \mathbf{\hat{y}}+p_z \mathbf{\hat{z}}=\vec{p}+p_z \mathbf{\hat{z}} $.
\end{table}

Since earlier work asserts that the pseudospin is not associated with an angular momentum \cite{1984DiVincenzo,2007Gusynin},  it is worth exploring why we are led to a different result. In a strictly 2D system angular momentum is defined only in a limited sense.   There is only one generator corresponding to (commuting) rotations about the direction normal to the plane \cite{2007Gusynin}, which is inconsistent with a 3+1 dimensional spin.  By confining the electrons, the graphene sheet produces an electronic Hamiltonian $\Ham \propto \mathbf{S}\cdot \mathbf{u}$ which lacks the full 3D rotational symmetry of the vacuum.  However, the (scalar) Hamiltonian is invariant under rotations of the system relative to the three coordinate axes, and the generators of these rotations, \ie the angular momentum operators, are still well defined. (Table~\ref{IsospinTable} juxtaposes the rotational properties of various spin-type Hamiltonians.) And unlike two-level systems that can be reduced to the spin-1/2 problem by analogy only, the honeycomb structure produces a spin variable specified in relation to a laboratory coordinate system. In other words, the direction of $\mathbf{u}$, like that of a magnetic field $\mathbf{B}$, can be related to `east', where no such relation appears in, say, the isospin problem.  Since the operator $R(\mathbf{n},\Phi)=\exp(-i\Phi \mathbf{S}\cdot \hat{\mathbf{n}}/\hbar)$ generates a rotation of the observable $\mathbf{S}$ by an angle $\Phi$ about the $\hat{\mathbf{n}}\equiv\mathbf{n}/|\mathbf{n}|$ axis, where $\hat{\mathbf{n}}$ is a direction in real space, $\mathbf{S}$ must be a real angular momentum.  See the Appendixes for a proof and further discussion.

\begin{table*}\begin{center}% * and center for 2 column I think, copying thermal imaging prl
\begin{tabular}{r||ccccc}
&neutrino\cite{1967Sakurai}&spin-orbit\cite{1977CohenT}&Zeeman\cite{1977CohenT}&honeycomb&deuteron\cite{1987Griffiths,1997Tomonaga}\\
\hline \hline
Hamiltonian $H$ $\propto \mathbf{S}\cdot\mathbf{Z}$&$\mathbf{S}\cdot \mathbf{p}$&$\mathbf{S}\cdot \mathbf{L}$&$\mathbf{S}\cdot \mathbf{B}$&$\mathbf{S}\cdot \mathbf{u}$&$\mathbf{S}^{(1)}\cdot\mathbf{S}^{(2)}$\\
 $\mathbf{S}$ name&spin&spin&spin&lattice spin&isospin\\
3D rotational symmetry& yes&yes & no & no & yes\\
$[H,\mathbf{J}]\propto$&0&0&$\mathbf{S}\times \mathbf{B}$&$\boldsymbol{\xi}_1$&0\\
$[\hat{\mathbf{n}}\cdot\mathbf{J},\mathbf{S}]/i\hbar$&$\mathbf{S}\times \hat{\mathbf{n}}$&$\mathbf{S}\times \hat{\mathbf{n}}$&$\mathbf{S}\times \hat{\mathbf{n}}$&$\mathbf{S}\times \hat{\mathbf{n}}$&0\\
$[\hat{\mathbf{n}}\cdot\mathbf{J},\mathbf{Z}]/i\hbar$ &$\mathbf{p}\times\hat{\mathbf{n}}$ &$\mathbf{L}\times\hat{\mathbf{n}}$& 0&$\boldsymbol{\xi}_2$&0\\
$\mathbf{S}= \text{angular momentum}$&yes&yes&yes&yes&no\\
\end{tabular}%oldschool $ v_F (\boldsymbol{\sigma}-\sigma_z \mathbf{\hat{z}})\times \mathbf{p}$
\caption{Hamiltonians with spin-like variables $\mathbf{S}$.  A general unit vector is designated by $\hat{\mathbf{n}}$, and the placeholders are $\boldsymbol{\xi}_1=\mathbf{S}\times (\mathbf{u}-\mathbf{p})+(\mathbf{S}\cdot\hat{\mathbf{a}}_n)(\hat{\mathbf{a}}_n\times\mathbf{p})$ and $\boldsymbol{\xi}_2= \mathbf{p}\times\hat{\mathbf{n}}-(\mathbf{p}\times\hat{\mathbf{n}})\cdot \hat{\mathbf{a}}_n \,\hat{\mathbf{a}}_n$.  While all Hamiltonians are rotational invariants in the sense that they are scalars, the Zeeman and honeycomb Hamiltonians do not represent rotationally invariant systems. Also note that $\mathbf{B}$, $\mathbf{u}$, and the isospin $\mathbf{S}$ are not spatial, quantum mechanical vectors $\mathbf{V}$, as they do not satisfy $[\hat{\mathbf{n}}\cdot\mathbf{J},\mathbf{V}]=i \hbar \mathbf{V}\times\hat{\mathbf{n}}$.\cite{1977CohenT} However, as the Zeeman example illustrates, this condition is not required for $\mathbf{S}$ to be an angular momentum.}
\label{IsospinTable}
\end{center}\end{table*}

The analogy between the vector $\mathbf{u}$ and a magnetic field $\mathbf{B}$ is helpful for understanding the close relationship between energy and angular momentum that follows from the $\boldsymbol{\sigma}$ operator's coverage of both qubits.  We compare the time evolution operator $U=\exp(-i \Ham t/\hbar)$ with the rotation operator $R(\mathbf{u},\Phi)$.  For the honeycomb Hamiltonian (\ref{eq:HamU}) these expressions are identical; time evolving the state is equivalent to rotating it about the  $\hat{\mathbf{u}}$ axis at a rate $d\Phi/dt= (E_+-E_-)/\hbar$.  The energies $E_\beta$ correspond to the two eigenvalues of the Hamiltonian $\Ham$,
\begin{equation}\label{eq:energy}
E_\beta = \beta \sqrt{v_F^2(p_x^2+p_y^2)+m'^2 v_F^4},
\end{equation}
where we have defined the band index $\beta=\pm 1$.  An equivalent relationship between time evolution and rotations appears when a magnetic dipole is placed in a magnetic field $\mathbf{B}$, where it is called Larmor precession.  The time evolution of the lattice spin operator (see Table~\ref{DiracTable}) is given by 
\begin{equation}\label{eq:Larmor}
\frac{d\mathbf{S}}{dt}=\frac{2  v_F}{\hbar}\mathbf{u}\times \mathbf{S},
\end{equation} which is exactly the equation of motion of a magnetic dipole in a magnetic field, with the substitution $2  v_F \mathbf{u}/\hbar \rightarrow -\gamma \mathbf{B}$ (here $\gamma$ is the gyromagnetic ratio).  Thus the mixing between helicity eigenstates produced by a mass term in the 2+1 dimensional Dirac Hamiltonian is formally identical to the mixing between spin-up and spin-down states produced by perpendicular magnetic field $\mathbf{B}_\perp$. In the general case of $\mathbf{u}$ not parallel to $\mathbf{S}$, Eq.~(\ref{eq:Larmor}) shows that the components of $\mathbf{S}$ perpendicular and parallel to the plane interconvert as a function of time. 

Relating pseudospin to angular momentum provides an intuitive explanation, the existence of which has long been suspected \cite{1998Ando}, for some properties of graphene and related materials.  For instance, in metallic nanotubes the backscattering of electrons is suppressed if the scattering potential has a range that is long compared to a lattice constant \cite{1998Ando,1999McEuen}.  Since such a potential will treat A and B sites identically, it is a scalar with respect to the lattice spin, and thus cannot create the lattice spin flip required to reverse the electron motion.  Conservation of  pseudospin or lattice spin also provides a way to understand  Klein tunneling in graphene \cite{2006KatsnelsonKlein}, and emphasizes the equivalence between the condensed matter and particle physics systems. 

As another example, interpreting the pseudospin as connected to a real angular momentum gives a satisfying picture of photon-mediated electron-hole pair production (or recombination) in graphene.  In this process an electron with pseudospin parallel to its momentum transitions between the negative energy valence band and the positive energy conduction band, absorbing (or emitting) a photon and flipping its pseudospin \cite{2008Kuzmenko,2008Nair}.  Both bands are derived from atomic $2P_z$ orbitals, and the electron spin does not flip, so the usual sources do not contribute to the photon's angular momentum. A full description \cite{2010Tree} of this process based on the Hamiltonian (\ref{eq:DiracP}) and the substitution $\mathbf{p}\rightarrow \mathbf{p}-q \mathbf{A}/ c$ (with the quantized vector potential $\mathbf{A}$ describing the photon) shows that the photon's polarization couples the initial and final pseudospin states.   As the photon's vector polarization is associated \cite{1967Sakurai} with its spin $\hbar$, the form of the transition matrix element requires that the pseudospin be associated with an angular momentum $\hbar/2$.

Thus all three components of the graphene pseudospin have straightforward connections with angular momentum: reversing an in-plane component creates a spin-1 particle, while the out-of-plane component clearly generates rotations.  These complementary examples and the interconversion described by the equation of motion (\ref{eq:Larmor}) make a compelling intuitive case for the pseudospin-angular momentum identity.

Spin states have been previously associated with specific types of sites in a lattice within the ``staggered'' formulation of lattice quantum chromodynamics \cite{1977Susskind}.  However, the hypercubic lattice chosen there is a convenient framework for discretizing the Dirac equation, not the basis of a more fundamental model from which the Dirac equation emerges naturally.  The hypercubic lattice cannot be viewed as a discretized space through which particles move, since necessary phase tuning in the lattice Hamiltonian \cite{1977Susskind,1991Zee} ruins its hopping interpretation for $D>1$.  Furthermore, the number of sublattices must be fixed artificially using foreknowledge of the Dirac equation.  These differences are not solely aesthetic; unlike the unphysical Hamiltonian contrived for the hypercubic lattice, the honeycomb Hamiltonian can be studied in optical lattices \cite{2011Soltan-Panahi} and natural materials such as graphene \cite{1984Semenoff,2008Kuzmenko,2008Nair}.

By generating an internal quantum number that reflects an energy scale much larger than the host particle's mass \cite{2008PDGcompositeness}, the graphene example invites connection to the intrinsic spin carried by the quarks and leptons.  However, it is not obvious that a lattice can be found that naturally produces the $3+1$ dimensional Dirac equation \cite{2008Creutz,2008Bedaque}, let alone the chiral properties of the full standard model \cite{1981NielsenHomotopy,2004Chandrasekharan}.  Thus there are two possibilities, depending on whether these spins have related origins. Either intrinsic spin is also the low-energy signature of nontrivial quantized space, or lattice spin represents a second, experimentally-accessible type of quantum mechanical angular momentum with no classical analogue.

\appendix%* add star to remove lettering
\section{Proof}\label{sec:proof}
By definition an operator $\mathbf{S}$ is a 3+1 dimensional angular momentum if its Cartesian components $S_i$ obey the commutation relations
\begin{equation}\label{eq:vector}
[S_i,S_j]=i \hbar \epsilon_{ijk} S_k,
\end{equation}
and it generates rotations in real space \cite{1977CohenT}. The Pauli matrices obey $[\sigma_i,\sigma_j]=2i \epsilon_{ijk} \sigma_k$, so $\mathbf{S}$ as defined by (\ref{eq:Lspin}) satisfies (\ref{eq:vector}).  The form (\ref{eq:Lspin}) also makes it clear, through the dependence on the $\hat{\mathbf{a}}_i$, that $\mathbf{S}$ is referenced to directions in real space.  To prove that the $S_i$ generate rotations we consider the Pauli matrices as components of $\mathbf{S}$, and rotate $\mathbf{S}$ by an angle $\Phi$ about an axis $\hat{\mathbf{n}}$ according to the classical relation \cite{2002Goldstein}
\begin{equation}
\mathbf{S}'=\mathbf{S} \cos \Phi + \hat{\mathbf{n}} (\hat{\mathbf{n}}\cdot \mathbf{S}) (1-\cos\Phi)+(\hat{\mathbf{n}}\times\mathbf{S})\sin\Phi,
\end{equation}
and compare the result with the $\mathbf{S}'$ found by performing the quantum rotation operation \cite{1977CohenT}, 
\begin{equation}
\mathbf{S}'=e^{i \Phi\mathbf{J}\cdot\hat{\mathbf{n}}/\hbar}\,\mathbf{S}\,e^{-i \Phi\mathbf{J}\cdot\hat{\mathbf{n}}/\hbar}.
\end{equation}
For total angular momentum of the form $\mathbf{J}=\mathbf{S}+\sum \mathbf{J}_{\text{other}}$, where $[\mathbf{S},\mathbf{J}_{\text{other}}]=0$, these two transformations give the same result, which completes the proof.

\section{Comparison with the Lorentz group in 2+1 D}\label{sec:Lorentz}
%At this point it is instructive reexamine the role of pseudospin $\boldsymbol{\sigma}$ from the $2+1$ dimensional perspective.  
The honeycomb pseudospin $\boldsymbol{\sigma}$ can be associated with two distinct algebras:
\begin{subequations}\label{eq:algebras}
\begin{align}
[S_i,S_j]&=i \hbar \epsilon_{ijk} S_k,&&(\{i,j,k\}\, \in \{1,2,3\}), \, &\text{and}\label{eq:SpinCom}\\
\{\gamma^\mu,\gamma^\nu\}&=2 g^{\mu\nu},&&(\{\mu,\nu\}\, \in \{0,1,2\}),&\label{eq:GammaAntiCom}
\end{align}
\end{subequations}
with respective representations
\begin{subequations}\label{eq:representations}
\begin{align}
S_i &=\frac{\hbar}{2} (\kappa \sigma_x, \sigma_y,\kappa \sigma_z), \quad \text{and}\label{eq:SiRep}\\
\gamma^\mu  &= (\sigma_z,i \sigma_x, i\kappa \sigma_y).\label{eq:GammaRep}
\end{align}
\end{subequations}
As discussed previously, the $S_i$ of (\ref{eq:SiRep}) generate rotations in 3 spatial dimensions by virtue of satisfying (\ref{eq:SpinCom}).  Similarly, the $\gamma^\mu$ of (\ref{eq:GammaRep}) are associated with two boosts and one rotation in $2+1$ dimensions by virtue of satisfying (\ref{eq:GammaAntiCom}).%of $\mathbf{S}$

The Dirac algebra  (\ref{eq:GammaAntiCom}) connects the gamma matrices $\gamma^\mu$ to the Minkowski metric $g^{\mu\nu}$, which here has signature $(+\, - \,-)$. In this representation \cite{1964Bjorken} the generators $L^{\mu\nu}$ of Lorentz transformations are given by $L^{\mu\nu}= \frac{i}{4}[\gamma^\mu,\gamma^\nu]$.  Explicit calculation gives 
\begin{subequations}
\begin{align}
K_1&=L^{01}=\frac{-i}{2}\sigma_y,\\
K_2&=L^{02}=\frac{i\kappa}{2}\sigma_x,\quad \text{and}\\
J_3&=L^{12}=\frac{\kappa}{2}\sigma_z,
\end{align}
\end{subequations}
where, as we will see, a $K_j$ generates a boost along the $j$ axis and $J_3$ generates the rotation mixing the $\{1,2\}$ coordinates.  

The form of the Dirac equation must be frame independent. A spinor $\ket{\chi}$ transforms according to $\ket{\chi'}=D_j \ket{\chi}$, with $D_j=\exp(-i \theta X_j)$ and the transformation generator $X_j \in \{K_1,K_2,J_3\}$.  If a spacetime three-vector, \eg $p^\mu$, has the corresponding transformation ${p'}^\mu =\Lambda^\mu_{\phantom{\mu}\nu}(j)\, p^\nu$, consistency then requires $\Lambda^\mu_{\phantom{\mu}\nu}(j) \gamma^\nu=D^{-1}_j\gamma^\mu D_j$.  Using this relation we find the $\Lambda^\mu_{\phantom{\mu}\nu}(j)$,
\begin{subequations}\label{eq:LorentzTransformations}
\begin{equation}\label{eq:LorentzBoosts}
\Lambda(1)= \begin{pmatrix}
\cosh \theta &\sinh \theta &0\\
\sinh \theta& \cosh \theta&0 \\
0 & 0&1 \end{pmatrix}, \,
\Lambda(2)= \begin{pmatrix}
\cosh \theta &0&\sinh \theta \\
0&1&0\\
\sinh \theta& 0&\cosh \theta \end{pmatrix},
\end{equation}
\begin{equation}
\text{and}\qquad
\Lambda(3)= \begin{pmatrix}
1 &0&0\\
0& \cos \theta& -\sin \theta \\
0 & \sin\theta &\cos \theta
\end{pmatrix},
\end{equation}
\end{subequations}%readily recognized as 
which are the two boosts and one rotation expected in $2+1$ dimensions.  The appearance of hyperbolic functions in the Lorentz boosts (\ref{eq:LorentzBoosts}) reveals the distinguishing role of the $i$'s which occur in the $\gamma^\mu$ representation (\ref{eq:GammaRep}) and are notably absent from the $S_i$ representation (\ref{eq:SiRep}).  

%Note also that while it is possible to define representations satisfying (\ref{eq:algebras}) that do not include $\kappa$'s, the forms (\ref{eq:representations}) are preferred.  Including the valley index $\kappa$ in the definitions of the $S_i$ and the $\gamma^\mu$ allows the Hamiltonian eigenstates (see \cite{2010Tree}) to be labeled by independent  $\kappa$, $\beta$, and $\mathbf{u}$.  If instead the $\kappa$'s were included in the definition of $\mathbf{u}$,  the expectation values for $\mathbf{S}$ and $\mathbf{u}$ would show undesirable behavior under change of $\kappa$. %previous penultimate paragraph reflecting bad definition of lattice spin

Note also that while it is possible to define representations satisfying (B1) that do not include $\kappa$'s, the forms (\ref{eq:representations}) are preferred.  If instead all of the $\kappa$'s were included in the definition of $\mathbf{u}$,  the expectation value of $\mathbf{S}$ for a Hamiltonian eigenstate would show undesirable behavior under change of $\kappa$.  More importantly, including the valley index $\kappa$ in the definitions of the $S_i$ and the $\gamma^\mu$ gives these operators the expected transformation properties under time reversal, which is implemented by taking $i\rightarrow -i$ and $\kappa\rightarrow - \kappa$.  This prescription can be deduced by noting that $\Ham$ is invariant under time reversal and comparing Eq.~(\ref{eq:Mmatrix}) and Eq.~(\ref{eq:linearMmatrix}) of the main text.

Thus the pseudospin Pauli matrices $\boldsymbol{\sigma}$ can be connected both to the generators of the Lorentz group in $2+1$ dimensions and the generators of the rotation group in $3+1$ dimensions.  The choice of algebra corresponds to considering the honeycomb lattice as a strictly two dimensional structure, or as a quasi-two dimensional structure embedded in three dimensional space.  Since experiments on graphene occur in three dimensional space containing inherently three dimensional objects (\eg the photon \cite{2010Tree}), this second perspective is sometimes unavoidable. %the key distinction is made by considering

\bibliography{graphene}% Produces the bibliography via BibTeX.

\begin{thebibliography}{31}
\expandafter\ifx\csname natexlab\endcsname\relax\def\natexlab#1{#1}\fi
\expandafter\ifx\csname bibnamefont\endcsname\relax
  \def\bibnamefont#1{#1}\fi
\expandafter\ifx\csname bibfnamefont\endcsname\relax
  \def\bibfnamefont#1{#1}\fi
\expandafter\ifx\csname citenamefont\endcsname\relax
  \def\citenamefont#1{#1}\fi
\expandafter\ifx\csname url\endcsname\relax
  \def\url#1{\texttt{#1}}\fi
\expandafter\ifx\csname urlprefix\endcsname\relax\def\urlprefix{URL }\fi
\providecommand{\bibinfo}[2]{#2}
\providecommand{\eprint}[2][]{\url{#2}}

\bibitem[{\citenamefont{Tomonaga}(1997)}]{1997Tomonaga}
\bibinfo{author}{\bibfnamefont{S.}~\bibnamefont{Tomonaga}},
  \emph{\bibinfo{title}{The story of spin}} (\bibinfo{publisher}{University of
  Chicago Press}, \bibinfo{address}{Chicago}, \bibinfo{year}{1997}).

\bibitem[{\citenamefont{Hagiwara et~al.}(2008)\citenamefont{Hagiwara, Hikasa,
  Tanabashi, and Group}}]{2008PDGcompositeness}
\bibinfo{author}{\bibfnamefont{K.}~\bibnamefont{Hagiwara}},
  \bibinfo{author}{\bibfnamefont{K.}~\bibnamefont{Hikasa}},
  \bibinfo{author}{\bibfnamefont{M.}~\bibnamefont{Tanabashi}},
  \bibnamefont{and} \bibinfo{author}{\bibfnamefont{P.~D.} \bibnamefont{Group}},
  \bibinfo{journal}{Physics Letters B} \textbf{\bibinfo{volume}{667}},
  \bibinfo{pages}{1209} (\bibinfo{year}{2008}).

\bibitem[{\citenamefont{Dirac}(1958)}]{1958Dirac}
\bibinfo{author}{\bibfnamefont{P.~A.~M.} \bibnamefont{Dirac}},
  \emph{\bibinfo{title}{The principles of quantum mechanics.}}
  (\bibinfo{publisher}{Clarendon Press}, \bibinfo{address}{Oxford},
  \bibinfo{year}{1958}).

\bibitem[{\citenamefont{DiVincenzo and Mele}(1984)}]{1984DiVincenzo}
\bibinfo{author}{\bibfnamefont{D.~P.} \bibnamefont{DiVincenzo}}
  \bibnamefont{and} \bibinfo{author}{\bibfnamefont{E.~J.} \bibnamefont{Mele}},
  \bibinfo{journal}{Physical Review B} \textbf{\bibinfo{volume}{29}},
  \bibinfo{pages}{1685} (\bibinfo{year}{1984}).

\bibitem[{\citenamefont{Semenoff}(1984)}]{1984Semenoff}
\bibinfo{author}{\bibfnamefont{G.~W.} \bibnamefont{Semenoff}},
  \bibinfo{journal}{Physical Review Letters} \textbf{\bibinfo{volume}{53}},
  \bibinfo{pages}{2449} (\bibinfo{year}{1984}).

\bibitem[{\citenamefont{Wallace}(1947)}]{1947Wallace}
\bibinfo{author}{\bibfnamefont{P.~R.} \bibnamefont{Wallace}},
  \bibinfo{journal}{Physical Review} \textbf{\bibinfo{volume}{71}},
  \bibinfo{pages}{622} (\bibinfo{year}{1947}).

\bibitem[{\citenamefont{Iijima}(1991)}]{1991Iijima}
\bibinfo{author}{\bibfnamefont{S.}~\bibnamefont{Iijima}},
  \bibinfo{journal}{Nature} \textbf{\bibinfo{volume}{354}}, \bibinfo{pages}{56}
  (\bibinfo{year}{1991}).

\bibitem[{\citenamefont{Novoselov et~al.}(2004)\citenamefont{Novoselov, Geim,
  Morozov, Jiang, Zhang, Dubonos, Grigorieva, and Firsov}}]{2004Novoselov}
\bibinfo{author}{\bibfnamefont{K.~S.} \bibnamefont{Novoselov}},
  \bibinfo{author}{\bibfnamefont{A.~K.} \bibnamefont{Geim}},
  \bibinfo{author}{\bibfnamefont{S.~V.} \bibnamefont{Morozov}},
  \bibinfo{author}{\bibfnamefont{D.}~\bibnamefont{Jiang}},
  \bibinfo{author}{\bibfnamefont{Y.}~\bibnamefont{Zhang}},
  \bibinfo{author}{\bibfnamefont{S.~V.} \bibnamefont{Dubonos}},
  \bibinfo{author}{\bibfnamefont{I.~V.} \bibnamefont{Grigorieva}},
  \bibnamefont{and} \bibinfo{author}{\bibfnamefont{A.~A.}
  \bibnamefont{Firsov}}, \bibinfo{journal}{Science}
  \textbf{\bibinfo{volume}{306}}, \bibinfo{pages}{666} (\bibinfo{year}{2004}).

\bibitem[{\citenamefont{Saito et~al.}(1998)\citenamefont{Saito, Dresselhaus,
  and Dresselhaus}}]{1998Saito}
\bibinfo{author}{\bibfnamefont{R.}~\bibnamefont{Saito}},
  \bibinfo{author}{\bibfnamefont{G.}~\bibnamefont{Dresselhaus}},
  \bibnamefont{and} \bibinfo{author}{\bibfnamefont{M.~S.}
  \bibnamefont{Dresselhaus}}, \emph{\bibinfo{title}{Physical properties of
  carbon nanotubes}} (\bibinfo{publisher}{Imperial College Press},
  \bibinfo{address}{London}, \bibinfo{year}{1998}).

\bibitem[{\citenamefont{Neto et~al.}(2009)\citenamefont{Neto, Guinea, Peres,
  Novoselov, and Geim}}]{2009NetoReview}
\bibinfo{author}{\bibfnamefont{A.~H.~C.} \bibnamefont{Neto}},
  \bibinfo{author}{\bibfnamefont{F.}~\bibnamefont{Guinea}},
  \bibinfo{author}{\bibfnamefont{N.~M.~R.} \bibnamefont{Peres}},
  \bibinfo{author}{\bibfnamefont{K.~S.} \bibnamefont{Novoselov}},
  \bibnamefont{and} \bibinfo{author}{\bibfnamefont{A.~K.} \bibnamefont{Geim}},
  \bibinfo{journal}{Reviews of Modern Physics} \textbf{\bibinfo{volume}{81}},
  \bibinfo{pages}{109} (\bibinfo{year}{2009}).

\bibitem[{\citenamefont{Katsnelson and Novoselov}(2007)}]{2007KatsnelsonBridge}
\bibinfo{author}{\bibfnamefont{M.~I.} \bibnamefont{Katsnelson}}
  \bibnamefont{and} \bibinfo{author}{\bibfnamefont{K.~S.}
  \bibnamefont{Novoselov}}, \bibinfo{journal}{Solid State Communications}
  \textbf{\bibinfo{volume}{143}}, \bibinfo{pages}{3} (\bibinfo{year}{2007}).

\bibitem[{\citenamefont{Katsnelson et~al.}(2006)\citenamefont{Katsnelson,
  Novoselov, and Geim}}]{2006KatsnelsonKlein}
\bibinfo{author}{\bibfnamefont{M.~I.} \bibnamefont{Katsnelson}},
  \bibinfo{author}{\bibfnamefont{K.~S.} \bibnamefont{Novoselov}},
  \bibnamefont{and} \bibinfo{author}{\bibfnamefont{A.~K.} \bibnamefont{Geim}},
  \bibinfo{journal}{Nature Physics} \textbf{\bibinfo{volume}{2}},
  \bibinfo{pages}{620} (\bibinfo{year}{2006}).

\bibitem[{\citenamefont{Bena and Montambaux}(2009)}]{2009Bena}
\bibinfo{author}{\bibfnamefont{C.}~\bibnamefont{Bena}} \bibnamefont{and}
  \bibinfo{author}{\bibfnamefont{G.}~\bibnamefont{Montambaux}},
  \bibinfo{journal}{New Journal of Physics} \textbf{\bibinfo{volume}{11}},
  \bibinfo{pages}{095003} (\bibinfo{year}{2009}).

\bibitem[{\citenamefont{Gusynin et~al.}(2007)\citenamefont{Gusynin, Sharapov,
  and Carbotte}}]{2007Gusynin}
\bibinfo{author}{\bibfnamefont{V.~P.} \bibnamefont{Gusynin}},
  \bibinfo{author}{\bibfnamefont{S.~G.} \bibnamefont{Sharapov}},
  \bibnamefont{and} \bibinfo{author}{\bibfnamefont{J.}~\bibnamefont{Carbotte}},
  \bibinfo{journal}{International Journal of Modern Physics B}
  \textbf{\bibinfo{volume}{21}}, \bibinfo{pages}{4611} (\bibinfo{year}{2007}).

\bibitem[{\citenamefont{Sakurai}(1967)}]{1967Sakurai}
\bibinfo{author}{\bibfnamefont{J.~J.} \bibnamefont{Sakurai}},
  \emph{\bibinfo{title}{Advanced quantum mechanics}}
  (\bibinfo{publisher}{Addison-Wesley Pub. Co.}, \bibinfo{address}{Reading,
  MA}, \bibinfo{year}{1967}).

\bibitem[{\citenamefont{Cohen-Tannoudji
  et~al.}(1977)\citenamefont{Cohen-Tannoudji, Diu, and Laloë}}]{1977CohenT}
\bibinfo{author}{\bibfnamefont{C.}~\bibnamefont{Cohen-Tannoudji}},
  \bibinfo{author}{\bibfnamefont{B.}~\bibnamefont{Diu}}, \bibnamefont{and}
  \bibinfo{author}{\bibfnamefont{F.}~\bibnamefont{Laloë}},
  \emph{\bibinfo{title}{Quantum mechanics}} (\bibinfo{publisher}{Wiley},
  \bibinfo{address}{New York}, \bibinfo{year}{1977}).

\bibitem[{\citenamefont{Griffiths}(1987)}]{1987Griffiths}
\bibinfo{author}{\bibfnamefont{D.~J.} \bibnamefont{Griffiths}},
  \emph{\bibinfo{title}{Introduction to elementary particles}}
  (\bibinfo{publisher}{Wiley}, \bibinfo{address}{New York},
  \bibinfo{year}{1987}).

\bibitem[{\citenamefont{Ando and Nakanishi}(1998)}]{1998Ando}
\bibinfo{author}{\bibfnamefont{T.}~\bibnamefont{Ando}} \bibnamefont{and}
  \bibinfo{author}{\bibfnamefont{T.}~\bibnamefont{Nakanishi}},
  \bibinfo{journal}{Journal of the Physical Society of Japan}
  \textbf{\bibinfo{volume}{67}}, \bibinfo{pages}{1704} (\bibinfo{year}{1998}).

\bibitem[{\citenamefont{McEuen et~al.}(1999)\citenamefont{McEuen, Bockrath,
  Cobden, Yoon, and Louie}}]{1999McEuen}
\bibinfo{author}{\bibfnamefont{P.~L.} \bibnamefont{McEuen}},
  \bibinfo{author}{\bibfnamefont{M.}~\bibnamefont{Bockrath}},
  \bibinfo{author}{\bibfnamefont{D.~H.} \bibnamefont{Cobden}},
  \bibinfo{author}{\bibfnamefont{Y.~G.} \bibnamefont{Yoon}}, \bibnamefont{and}
  \bibinfo{author}{\bibfnamefont{S.~G.} \bibnamefont{Louie}},
  \bibinfo{journal}{Physical Review Letters} \textbf{\bibinfo{volume}{83}},
  \bibinfo{pages}{5098} (\bibinfo{year}{1999}).

\bibitem[{\citenamefont{Kuzmenko et~al.}(2008)\citenamefont{Kuzmenko, van
  Heumen, Carbone, and van~der Marel}}]{2008Kuzmenko}
\bibinfo{author}{\bibfnamefont{A.~B.} \bibnamefont{Kuzmenko}},
  \bibinfo{author}{\bibfnamefont{E.}~\bibnamefont{van Heumen}},
  \bibinfo{author}{\bibfnamefont{F.}~\bibnamefont{Carbone}}, \bibnamefont{and}
  \bibinfo{author}{\bibfnamefont{D.}~\bibnamefont{van~der Marel}},
  \bibinfo{journal}{Physical Review Letters} \textbf{\bibinfo{volume}{100}},
  \bibinfo{pages}{117401} (\bibinfo{year}{2008}).

\bibitem[{\citenamefont{Nair et~al.}(2008)\citenamefont{Nair, Blake,
  Grigorenko, Novoselov, Booth, Stauber, Peres, and Geim}}]{2008Nair}
\bibinfo{author}{\bibfnamefont{R.~R.} \bibnamefont{Nair}},
  \bibinfo{author}{\bibfnamefont{P.}~\bibnamefont{Blake}},
  \bibinfo{author}{\bibfnamefont{A.~N.} \bibnamefont{Grigorenko}},
  \bibinfo{author}{\bibfnamefont{K.~S.} \bibnamefont{Novoselov}},
  \bibinfo{author}{\bibfnamefont{T.~J.} \bibnamefont{Booth}},
  \bibinfo{author}{\bibfnamefont{T.}~\bibnamefont{Stauber}},
  \bibinfo{author}{\bibfnamefont{N.~M.~R.} \bibnamefont{Peres}},
  \bibnamefont{and} \bibinfo{author}{\bibfnamefont{A.~K.} \bibnamefont{Geim}},
  \bibinfo{journal}{Science} \textbf{\bibinfo{volume}{320}},
  \bibinfo{pages}{1308} (\bibinfo{year}{2008}).

\bibitem[{\citenamefont{Mecklenburg et~al.}(2010)\citenamefont{Mecklenburg,
  Woo, and Regan}}]{2010Tree}
\bibinfo{author}{\bibfnamefont{M.}~\bibnamefont{Mecklenburg}},
  \bibinfo{author}{\bibfnamefont{J.}~\bibnamefont{Woo}}, \bibnamefont{and}
  \bibinfo{author}{\bibfnamefont{B.~C.} \bibnamefont{Regan}},
  \bibinfo{journal}{Physical Review B} \textbf{\bibinfo{volume}{81}},
  \bibinfo{pages}{245401} (\bibinfo{year}{2010}).

\bibitem[{\citenamefont{Susskind}(1977)}]{1977Susskind}
\bibinfo{author}{\bibfnamefont{L.}~\bibnamefont{Susskind}},
  \bibinfo{journal}{Physical Review D} \textbf{\bibinfo{volume}{16}},
  \bibinfo{pages}{3031} (\bibinfo{year}{1977}).

\bibitem[{\citenamefont{Zee}(1991)}]{1991Zee}
\bibinfo{author}{\bibfnamefont{A.}~\bibnamefont{Zee}}, in
  \emph{\bibinfo{booktitle}{M.A.B. Beg Memorial Volume}}, edited by
  \bibinfo{editor}{\bibfnamefont{A.}~\bibnamefont{Ali}} \bibnamefont{and}
  \bibinfo{editor}{\bibfnamefont{P.}~\bibnamefont{Hoodbhoy}}
  (\bibinfo{publisher}{World Scientific}, \bibinfo{address}{Singapore; River
  Edge, NJ}, \bibinfo{year}{1991}), pp. \bibinfo{pages}{129--140}.

\bibitem[{\citenamefont{Soltan-Panahi et~al.}(2011)\citenamefont{Soltan-Panahi,
  Struck, Hauke, Bick, Plenkers, Meineke, Becker, Windpassinger, Lewenstein,
  and Sengstock}}]{2011Soltan-Panahi}
\bibinfo{author}{\bibfnamefont{P.}~\bibnamefont{Soltan-Panahi}},
  \bibinfo{author}{\bibfnamefont{J.}~\bibnamefont{Struck}},
  \bibinfo{author}{\bibfnamefont{P.}~\bibnamefont{Hauke}},
  \bibinfo{author}{\bibfnamefont{A.}~\bibnamefont{Bick}},
  \bibinfo{author}{\bibfnamefont{W.}~\bibnamefont{Plenkers}},
  \bibinfo{author}{\bibfnamefont{G.}~\bibnamefont{Meineke}},
  \bibinfo{author}{\bibfnamefont{C.}~\bibnamefont{Becker}},
  \bibinfo{author}{\bibfnamefont{P.}~\bibnamefont{Windpassinger}},
  \bibinfo{author}{\bibfnamefont{M.}~\bibnamefont{Lewenstein}},
  \bibnamefont{and}
  \bibinfo{author}{\bibfnamefont{K.}~\bibnamefont{Sengstock}},
  \bibinfo{journal}{arXiv:1005.1276v1 [Nature Phys. (in press)]}
  (\bibinfo{year}{2011}).

\bibitem[{\citenamefont{Creutz}(2008)}]{2008Creutz}
\bibinfo{author}{\bibfnamefont{M.}~\bibnamefont{Creutz}},
  \bibinfo{journal}{Journal of High Energy Physics} \bibinfo{issue}{04}
  (\bibinfo{year}{2008}) \bibinfo{pages}{017}.

\bibitem[{\citenamefont{Bedaque et~al.}(2008)\citenamefont{Bedaque, Buchoff,
  Tiburzi, and Walker-Loud}}]{2008Bedaque}
\bibinfo{author}{\bibfnamefont{P.~F.} \bibnamefont{Bedaque}},
  \bibinfo{author}{\bibfnamefont{M.~I.} \bibnamefont{Buchoff}},
  \bibinfo{author}{\bibfnamefont{B.~C.} \bibnamefont{Tiburzi}},
  \bibnamefont{and}
  \bibinfo{author}{\bibfnamefont{A.}~\bibnamefont{Walker-Loud}},
  \bibinfo{journal}{Physical Review D} \textbf{\bibinfo{volume}{78}},
  \bibinfo{pages}{017502} (\bibinfo{year}{2008}).

\bibitem[{\citenamefont{Nielsen and Ninomiya}(1981)}]{1981NielsenHomotopy}
\bibinfo{author}{\bibfnamefont{H.~B.} \bibnamefont{Nielsen}} \bibnamefont{and}
  \bibinfo{author}{\bibfnamefont{M.}~\bibnamefont{Ninomiya}},
  \bibinfo{journal}{Nuclear Physics B} \textbf{\bibinfo{volume}{185}},
  \bibinfo{pages}{20} (\bibinfo{year}{1981}).

\bibitem[{\citenamefont{Chandrasekharan and Wiese}(2004)}]{2004Chandrasekharan}
\bibinfo{author}{\bibfnamefont{S.}~\bibnamefont{Chandrasekharan}}
  \bibnamefont{and} \bibinfo{author}{\bibfnamefont{U.~J.} \bibnamefont{Wiese}},
  \bibinfo{journal}{Progress in Particle and Nuclear Physics, Vol 53, No 2}
  \textbf{\bibinfo{volume}{53}}, \bibinfo{pages}{373} (\bibinfo{year}{2004}).

\bibitem[{\citenamefont{Goldstein et~al.}(2002)\citenamefont{Goldstein, Poole,
  and Safko}}]{2002Goldstein}
\bibinfo{author}{\bibfnamefont{H.}~\bibnamefont{Goldstein}},
  \bibinfo{author}{\bibfnamefont{C.}~\bibnamefont{Poole}}, \bibnamefont{and}
  \bibinfo{author}{\bibfnamefont{J.}~\bibnamefont{Safko}},
  \emph{\bibinfo{title}{Classical mechanics}} (\bibinfo{publisher}{Addison
  Wesley}, \bibinfo{address}{San Francisco}, \bibinfo{year}{2002}).

\bibitem[{\citenamefont{Bjorken and Drell}(1964)}]{1964Bjorken}
\bibinfo{author}{\bibfnamefont{J.~D.} \bibnamefont{Bjorken}} \bibnamefont{and}
  \bibinfo{author}{\bibfnamefont{S.~D.} \bibnamefont{Drell}},
  \emph{\bibinfo{title}{Relativistic quantum mechanics}}
  (\bibinfo{publisher}{McGraw-Hill}, \bibinfo{address}{New York},
  \bibinfo{year}{1964}).

\end{thebibliography}
\newpage

%
%\begin{acknowledgments}

%\end{acknowledgments}
%\newpage %Just because of unusual number of tables stacked at end

\end{document}